\documentclass[twocolumn,nopacs,preprintnumbers,amsmath,amssymb]{revtex4}

\usepackage{graphicx}
\usepackage{dcolumn}
\usepackage{bm}

\begin{document}

\preprint{APS/123-QED}
\title{Fluctuation Amplitude of a Trapped Rigid Sphere\\
Immersed in a Near-Critical Binary Fluid Mixture\\
within the Regime of the Gaussian Model}

\author{Youhei Fujitani}
 \email{youhei@appi.keio.ac.jp}
\affiliation{School of Fundamental Science and Technology,
Keio University, 
Yokohama 223-8522, Japan}

\date{\today}

\begin{abstract}
The position of a colloidal particle trapped in an external field
thermally fluctuates at equilibrium.
As is well known, the ambient fluid is not a simple heat bath and the particle mass
appears to increase, which influences the mean square velocity of the particle.
In this study, we suppose that the particle is surrounded by a binary fluid mixture
in the homogeneous phase near, but not too close to, the critical point.
Usually, one component is preferably attracted by the particle surface, and
the resultant adsorption layer becomes significant because of the near-criticality.
When the particle fluctuates in this situation, 
its mean square displacement should also be influenced by the ambient fluid
because the adsorption layer does not follow
the particle motion totally.   We calculate the influence in a simple case, where 
a rigid spherical particle fluctuates with a small amplitude and 
its surface attracts one component weakly.  We utilize
the hydrodynamics in the limit of no dissipation to examine the contribution 
from the ambient mixture to the equal-time correlation, which is shown to be reduced by
an additional stress, including osmotic pressure. 
\end{abstract}

\maketitle
\section{\label{sec:intro}Introduction}
Suppose a colloidal particle, about $0.1$--$1\ \mu$m in size, trapped by optical tweezers.
The particle position fluctuates at equilibrium, which is experimentally
observed with high resolutions of space ($<1$ nm) and time ($< 1 \ \mu$s) \cite{lukic,nat,natp,grim}.
For simplicity, we assume the one-dimensional motion of a rigid sphere (mass $m$) trapped in a
harmonic potential (natural angular frequency $\omega_0$) to consider the fluctuation.
The positional deviation of the particle center from the potential bottom is denoted by 
$\zeta$, while  the equilibrium average at the temperature $T$ is indicated by
$\langle\cdots\rangle$.   
The equal-time correlations can be calculated as
\begin{equation}
\langle \zeta^2 \rangle ={k_{\rm B}T\over m\omega_0^2}\quad {\rm and}\quad 
\langle \left({d\zeta\over dt}\right)^2 \rangle ={k_{\rm B}T\over m+m_{\rm ind}}
\ .\label{eqn:equaltime}\end{equation} 
Here, $k_{\rm B}$ denotes the Boltzmann constant, while
$m_{\rm ind}$ denotes the induced mass, which equals half the mass of the displaced
fluid \cite{lamb}.  The ambient fluid cannot be regarded as a heat bath; the particle mass appears to increase
because the particle motion causes flow.
The induced mass can be calculated in terms of the hydrodynamics for a one-component fluid.  \\

\begin{figure}
\includegraphics[width=6cm]{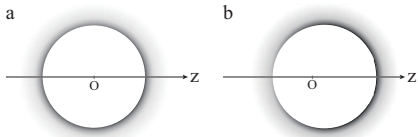}
\caption{\label{fig:twosphere}  
Cross section of a spherical particle immersed in a near-critical binary
fluid mixture is schematically drawn.  The gray scale  reflects
the mass-density difference between the two components; the shaded region represents the adsorption layer. 
(a) The particle  is fixed with its center being located at the origin $O$.
The adsorption layer is concentric with the particle contour in this figure; the mass-density difference is 
isotropic as viewed from the particle center.  (b) At a transient position of the particle in its motion along the $z$-axis,
the adsorption layer is deformed to become anisotropic because the
ambient fluid is not shifted translationally.  }
\end{figure}

In Eq.~(\ref{eqn:equaltime}), the mean square displacement
$\langle \zeta^2 \rangle$ is not influenced by
the fluid motion.  It is expected not to be the case, however, if the ambient fluid
is a near-critical binary fluid mixture.
The reason is as follows.   
It is usual that one component is preferably attracted by the particle surface \cite{Cahn}.
Suppose that the correlation length of the mixture is several nanometers, which is
much smaller than the particle radius, and that
the ambient fluid is in the homogeneous phase.
The adsorption layer, where the preferred component is more concentrated, appears around the
particle and has a thickness comparable to the correlation length \cite{binder,holyst}.
The adsorption layer is isotropic when the particle is fixed, 
as is drawn schematically in Fig.~\ref{fig:twosphere}(a).
There, although the resultant gradient of the mass-density difference 
between the two components generates osmotic pressure,  
the total force exerted on the particle by the ambient fluid vanishes.
When the particle moves rather rapidly, the adsorption layer cannot follow the motion
and is deformed, as shown in Fig.~\ref{fig:twosphere}(b).
Then, the additional force due to the gradient of the mass-density difference 
can be exerted on the particle in total.  It is thus possible that the ambient fluid
influences the mean square displacement, unlike in Eq.~(\ref{eqn:equaltime}). 
In this paper, we calculate this influence in a simple case. 
We assume the particle to be neutral electrically and no ion to be involved in the mixture. \\

We refer to some backgrounds.
In this paragraph, we mention how the induced mass appears in Eq.~(\ref{eqn:equaltime}), where
the ambient near-criticality combined with
the preferential attraction is not assumed.
The evolution of the particle position $\zeta$ 
with respect to the time $t$ can be described by the Langevin equation \cite{kubo,grim}
\begin{equation}
m{d^2\over dt^2}\zeta(t)= -m\omega_0^2\zeta(t)+F_{\rm hy}(t)+F_{\rm th}(t)
\ ,\label{eqn:lange}\end{equation}
where $F_{\rm hy}$ and $F_{\rm th}$
denote the force exerted by the ambient fluid and thermal noise, respectively.  
No particle rotation is assumed.
The force $F_{\rm hy}$ consists of the term $-m_{\rm ind} d^2\zeta/(dt^2)$, the term
causing the Stokes law, and the term having the memory effect.  The latter two terms 
involve the viscosity.   To calculate the equal-time correlation, we need not
take into account the dissipation and noise, irrespective of the real dynamics.
The reversible part of the dynamics,
\begin{equation}
\left(m+m_{\rm ind}\right){d^2\over dt^2} \zeta(t)= -m\omega_0^2\zeta\ ,\label{eqn:lange2}
\end{equation}  describes the oscillation about the equilibrium point.   From this, we find 
\begin{equation}
{1\over 2} m\omega_0^2 \zeta^2+ {1\over 2} (m+m_{\rm ind}) \left({d\zeta\over dt}\right)^2
\label{eqn:pote}\end{equation}
to give the total energy \cite{const}.  Thus, we arrive at Eq.~(\ref{eqn:equaltime}) with the aid of the equipartition theorem. \\

Similarly, 
we can use the reversible part of the hydrodynamics
to calculate the mean square displacement influenced by the deformed adsorption layer,
considering that
the particle radius is much larger than the correlation length.
To formulate the hydrodynamics, we use the Gaussian free-energy functional
by assuming the mixture to be in the homogeneous phase near, but not too close to,
the critical point.   In Appendix \ref{app:renorm},  
we estimate the temperature range validating the Gaussian model when the mixture has the critical composition
far from the surface.   
The hydrodynamics formulated from the coarse-grained free-energy functional can be found in the model H, where
the dissipation is assumed for
studying the relaxation of the two-time correlation in a near-critical fluid \cite{Kawa, Hohenberg, Onukibook}.  
In the present study, not interested in the critical slowing down, we 
utilize only the reversible part for calculating the equal-time correlation
although the real dynamics is dissipative \cite{ofk,furu,wetraft,wetdrop,visc,yabu}.
\\

The preferential attraction is assumed here to be caused by a short-range interaction.
It is thus represented by the additional free-energy functional determined
by the mass-density difference immediately near the particle surface.
Similar problems were studied in terms of the three-dimensional Ising model in a finite lattice 
\cite{binder, bray, diehl86, burk, diehl97}.
 If the coupling between neighboring spins is much stronger at the lattice boundary than in the bulk,  the 
bulk suffers the extraordinary second-order transition in the presence of the spontaneously ordered
surface when there is no external field.  The normal transition occurs when the surface is ordered by a surface field imposed
externally.  These two transitions share the same universal properties, and the latter can also be observed
in a binary fluid mixture in a container with the preferential attraction.  
These properties appear beyond the regime of the Gaussian model.
In the present study, to show that the deformed adsorption layer can generate the
additional force exerted on the particle, we consider a simple preferential attraction; 
the additional free-energy density is assumed to be a linear function \cite{ofk,furu,wetraft,wetdrop,visc,yabu}.
 \\

The present author has recently studied the mean square amplitude of the
shape fluctuation of a fluid membrane immersed in a near-critical binary fluid mixture
within the regime of the Gaussian model, and showed
that the deformed adsorption layer tends to suppress the amplitude \cite{pre}.
Although the particle itself is not deformed during its motion, unlike the membrane, 
the suppression effect of the deformed adsorption layer is also expected for the  
fluctuation amplitude of the particle position.
Our formulation and the outline of the calculation procedure are given  
in Sect.~\ref{sec:form}; the formulation for the mixture is the same as that used in Ref.~\onlinecite{pre}.  
We show the result in Sect.~\ref{sec:results}, and then describe the calculation procedure
in Sect.~\ref{sec:calc}.  
Our calculation is performed within the linear approximation of the fluctuation amplitude
and on the assumption of weak preferential attraction.
Equation (\ref{eqn:perpend})
represents the small oscillation of the particle about the equilibrium point.  
From this equation, we can find the potential energy modified by
the deformed adsorption layer.  Applying the equipartition theorem, we find 
the mean square displacement to be given by Eq.~(\ref{eqn:equi}), instead of 
the first equation of Eq.~(\ref{eqn:equaltime}).
The procedure becomes rather involved because a boundary layer appears in the limit of
vanishing interdiffusion.  We discuss the result by using
typical values of material constants and give some outlook in Sect.~\ref{sec:res}.  \\

\section{\label{sec:form} Formulation}
For the mass-density difference between the two components, we write $\varphi$, which
depends on the position $\bm{ r}$ in the mixture.  
As mentioned in Sect.~\ref{sec:intro}, we assume 
the $\varphi$-dependent part of the free-energy functional to be
\begin{eqnarray}&&
\int_{C^{\rm e}} d\bm{ r}\ 
\left[ f(\varphi(\bm{ r}))+{1\over 2}M
\left\vert\nabla\varphi(\bm{ r})\right\vert^2\right]
\nonumber\\ &&\qquad +\int_{\partial C }dS\ 
f_{{\rm s}}(\varphi(\bm{ r}))
\ ,\label{eqn:glw}\end{eqnarray}
where the coefficient $M$ is a positive constant.
The first term is the volume integral over the 
mixture region ($C^{\rm e}$), 
while the second is the surface integral
over the particle surface ($\partial C$).  
We assume the Gaussian model and the
preferential attraction caused by the surface field; 
$f$ is a quadratic function and
$f_{\rm s}$ is a linear function.
We write $h$ for the surface field, which is a constant defined as
$-f_{\rm s}'=-df_{\rm s}(\varphi)/(d\varphi)$.
Hereafter, the prime indicates the derivative with respect to the
variable, and the double prime
indicates the second derivative.\\

If we regard the mixture as a simple bath,
we can calculate the mean square displacement directly from Eq.~(\ref{eqn:glw})
without using the hydrodynamics, as shown in Appendix \ref{app:simple}.
Then, the chemical potential conjugate to $\varphi$ is homogeneous and constant, and
the profile of $\varphi$ is shifted translationally to follow the particle motion, {\it i.e.\/}, 
the adsorption layer moves with its distribution around the moving particle
being kept the same as shown in Fig.~\ref{fig:twosphere}(a).
Thus, in this improper calculation, the total force exerted on the particle remains unchanged 
from the one considered in Eq.~(\ref{eqn:lange2}).
To properly consider  the change in the chemical potential correlated with the particle motion,
we use the reversible dynamics based on Eq.~(\ref{eqn:glw})
to study the small oscillation of a particle about the equilibrium point,
as mentioned in Sect.~\ref{sec:intro}.  We write $\bm{ v}$
for the velocity field in the mixture.  
Assuming the mass density of the mixture
$\rho$ to be a constant, we have 
\begin{equation}
\nabla\cdot \bm{ v}(\bm{  r},t)=0\ .\label{eqn:incomp}
\end{equation} The chemical potential conjugate to $\varphi$ is given by
\begin{equation}
\mu(\bm{  r},t)=f'(\varphi(\bm{  r},t))
-M\Delta\varphi(\bm{  r},t) \ ,
\label{eqn:hatmudef}
\end{equation}
while the local equilibrium at the interface gives \cite{Cahn,ofk,jans}  
\begin{equation}
 M\bm{ n}\cdot \nabla 
\varphi =-h \quad {\rm at}\ \partial C
\label{eqn:phisurface}\ .\end{equation}
Here, $\bm{n}$ denotes the unit vector which is
normal to the particle surface and is directed outwards. 
We need not assume the viscosity to calculate the equal-time correlation, and have
\begin{equation}
\rho{\partial \bm{v}\over \partial t}
=-\nabla {p}-\varphi \nabla\mu \ ,
\label{eqn:3Ddyn}\end{equation}
where the convective term is neglected in anticipation of the
later linear approximation. 
The scalar ${p}$ originates from the $\rho$-dependent part of the free energy, 
but can be regarded as
dependent on $\bm{ r}$ and $t$ irrespective of the local state in the incompressible fluid.
The boundary-layer problem appears in the limit of zero viscosity.
We can deal with this problem by  
assuming the slip boundary condition at the particle surface.
There, the tangential components
of the velocity need not be
continuous, while the normal component is continuous. 
The stress exerted on the particle is evaluated immediately outside the boundary layer.\\

The diffusive flux between the two components
is proportional to the gradient of $\mu$.
The mass conservation of each component leads to   
\begin{equation}
{\partial\varphi\over\partial t}
=-\bm{ v}\cdot \nabla\varphi+ L\Delta \mu\ ,
\label{eqn:phidyn}\end{equation}
where the Onsager coefficient $L$ is assumed to be a positive constant. 
The diffusion flux cannot pass across the particle surface, which 
leads to
\begin{equation}
\bm{ n}\cdot L\nabla\mu=0\quad {\rm at}\ \partial C\ .
\label{eqn:musurface}\end{equation}
The diffusion should not be involved in the reversible dynamics, and
we should take the limit of $L\to 0+$. 
(Here, $0+$ means that the limit $L\to 0$ is taken with $L>0$ maintained.)
However, care should be taken because 
$L$ is associated with the highest-order derivative in Eq.~(\ref{eqn:phidyn}) \cite{bender}.
If $h$ does not vanish,
the limit $L\to 0+$ causes another boundary-layer problem, which is unfamiliar unlike the
problem in the limit of zero viscosity. 
Hence, we do not take the limit of $L\to 0+$ until we are able to examine
the influence of the boundary layer.  Details are shown in Sects.~\ref{sec:nond} and \ref{sec:sol}.  \\

Far from the particle, 
the mixture is assumed to be static and in the homogeneous phase, {\it i.e.\/}, $\bm{v}$ vanishes
and $\varphi$ is constant.  There, 
$\mu$ and ${p}$ are also constants,
considering Eqs.~(\ref{eqn:hatmudef}) and (\ref{eqn:3Ddyn}).
We write $\varphi_\infty$, $\mu^{(0)}\equiv f'(\varphi_\infty)$, 
and ${p}^{(0)}$ for the constant values of $\varphi$, $\mu$, and ${p}$, respectively.  
We assume the Gaussian model 
\begin{equation}
f(\varphi)={a\over 2} \left(\varphi-\varphi_\infty\right)^2
+\mu^{(0)}\left(\varphi-\varphi_\infty\right)
\ .\label{eqn:gauss}\end{equation}
Here, $a$ is a positive constant proportional to $T-T_{\rm c}$, where $T_{\rm c}$
denotes the critical temperature.
The correlation length far from the particle, denoted by $\xi_{\rm c}$, 
is given by $\sqrt{M/a}$.  
We nondimensionalize it to define  $s_{\rm c}$ as  
\begin{equation}
s_{\rm c} \equiv {\xi_{{\rm c}} \over r_0}=
{1\over r_0}\sqrt{{M\over a}}\ ,\label{eqn:zetac}
\end{equation}
where $r_0$ denotes the particle radius.  \\

\begin{figure}
\includegraphics[width=6cm]{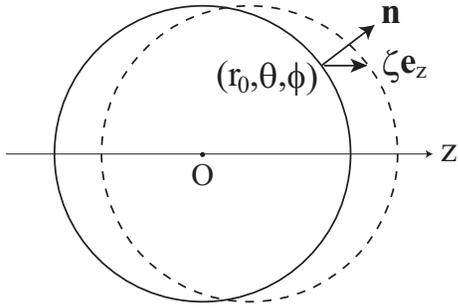}
\caption{\label{fig:sphere} 
We draw a cross section including the $z$-axis.  The bottom of the harmonic potential
is fixed at the origin $O$.  
 In the unperturbed state, 
the particle center is fixed at $O$ and the particle surface is represented by
the solid circle.   A point on the surface, with its spherical coordinates given
by $(r_0,\theta,\phi)$, moves when the particle is shifted; $r_0$ denotes the particle radius. 
A transient particle surface is
drawn with the dashed circle, and the displacement vector of the point is given by $\zeta \bm{e}_z$.
The unit normal vector $\bm{n}$ at the point remains unchanged in the particle motion. 
The dimensionless radial length is defined as $s\equiv r/r_0$.}
\end{figure}

We take the spherical polar coordinates
$(r,\theta,\phi)$, locating the origin at the bottom of the harmonic potential imposed externally.  
The particle motion is assumed to be along the $z$-axis (polar axis); 
$\bm{e}_z$ denotes the unit vector in the $z$-direction (Fig.~\ref{fig:sphere}).  We write $\zeta(t)$
for  the $z$-coordinate of the particle center.
The total force exerted on the particle by the mixture is along the $z$-axis;
its $z$-component is denoted by ${\cal F}_z(t)$.
The reversible dynamics of the particle is given by 
\begin{equation}
m{d^2  \over dt^2}\zeta(t)= {{\cal F}}_z(t) -m\omega_0^2 \zeta(t)
\label{eqn:spring}\end{equation}
instead of Eq.~(\ref{eqn:lange2}).  We should calculate $ {{\cal F}}_z$
from the fields of the mixture.  
Let us introduce the dimensionless parameter
$\varepsilon$ so that we have
\begin{equation}
\zeta(t) =\varepsilon \zeta^{(1)}(t)\ .
\end{equation}
Assuming $|\varepsilon|$ to be small, we calculate 
${\cal F}_z$ up to the order of $\varepsilon$ in Sect.~\ref{sec:sol}. \\

The equilibrium state occurring when the particle center is fixed at the origin
is regarded as the reference state or the unperturbed state, where 
$\mu$ is homogeneous over a mixture region and so is ${p}$ 
because of Eq.~(\ref{eqn:3Ddyn}) \cite{ofk}. They are respectively given by the constants
$\mu^{(0)}$ and ${p}^{(0)}$.  The superscript $^{(0)}$ is used to indicate the field in the unperturbed state.
Because of the symmetry of the unperturbed state,
$\varphi^{(0)}$ depends only on $r\equiv |\bm{r}|$.  As shown in Sect.~\ref{sec:unp},
we find it to be given by 
\begin{equation}
\varphi^{(0)}(r)
=\varphi_{\infty} + 
{hr_0e^{(1-s)/s_{\rm c}}\over Ms(1+s_{\rm c}^{-1})}
\ ,\label{eqn:phizero}\end{equation} 
where $s\equiv r/r_0$ is the dimensionless
radial length. This result is obtained in Ref.~\onlinecite{holyst} and is
used in Ref.~\onlinecite{ofk}.   Up to the order of $\varepsilon$, 
we expand the fields as
\begin{eqnarray}
& &\varphi(\bm{ r},t)=\varphi^{(0)}(r)+\varepsilon 
\varphi^{(1)}(\bm{ r},t)\ , \nonumber\\
&& \ \mu(\bm{ r},t)=\mu^{(0)}+\varepsilon \mu^{(1)}(\bm{ r},t)\ ,
\nonumber\\
& &{p}(\bm{ r},t)={p}^{(0)}+\varepsilon {p}^{(1)}(\bm{ r},t)\ ,\nonumber\\
&&\ {\rm and }\ \bm{ v}(\bm{ r},t)=\varepsilon \bm{ v}^{(1)}(\bm{ r},t)
\ .\label{eqn:perexp}\end{eqnarray}
The field with the
superscript $^{(1)}$ is defined so that it becomes
proportional to $\varepsilon$ after being multiplied by
$\varepsilon$.  The fields with the superscript $^{(1)}$ in Eq.~(\ref{eqn:perexp})
vanish far from the particle.  \\

We add an overtilde to the Fourier transform with respect to $t$, e.g.,
\begin{equation}
{\tilde p}^{(1)}(\bm{r},\omega)
={1\over 2\pi} \int_{-\infty}^\infty dt\  
p^{(1)}(\bm{ r}, t)
e^{i\omega t}\ .\label{eqn:fourtime}
\end{equation} 
Let $(r,\theta,\phi)$ be the components of the positional vector $\bm{r}$ in the mixture.
Using a spherical harmonics $Y_{10}(\theta)=\sqrt{3/(4\pi)}\cos{\theta}$, because of
the symmetry of the state, we can assume
\begin{equation}
{\tilde p}^{(1)}(\bm{r},\omega)=p_{10}(r,\omega)Y_{10}(\theta)\ ,
\label{eqn:pexp10}\end{equation}
whereby the coefficient function $p_{10}$ is defined.
Similarly, we introduce the coefficient functions for 
the Fourier transforms of the other mixture fields at the order of $\varepsilon$.
From Eqs.~(\ref{eqn:incomp}), (\ref{eqn:hatmudef}), (\ref{eqn:3Ddyn}), and  (\ref{eqn:phidyn}),
we can obtain a set of simultaneous equations with respect to these coefficient functions.
The equations are given by
Eqs.~(\ref{eqn:eqR10}), (\ref{eqn:subst010}), and (\ref{eqn:calG})
in Sect.~\ref{sec:calc}; we assume 
the preferential attraction to be weak to solve them approximately.   
Before describing our calculation procedure,
we show the result in the next section. \\

\section{\label{sec:results} Result}
We define $\Xi(s)$ so that ${\varphi^{(0)}}'(r)$ equals $r_0 \Xi(s) {\varphi^{(0)}}''(r_0)$, {\it i.e.\/}, 
\begin{equation}
\Xi(s)= -{s+s_{\rm c}\over s^2\kappa\left(1+s_{\rm c}\right) }e^{(1-s)/s_{\rm c}}
\ ,\label{eqn:Xidef}\end{equation}
where we use
\begin{equation}
\kappa\equiv 2+ {1\over s_{\rm c}\left(1+s_{\rm c}\right) }
\ .\label{eqn:kappadef}\end{equation}
We have ${\varphi^{(0)}}''(r_0)=h\kappa/(Mr_0)$ because of
Eq.~(\ref{eqn:phizero}).
Let us define the positive dimensionless parameter $\Lambda$ so that we have
\begin{equation}
\Lambda^2\equiv {h^2\kappa\over\rho\omega^2 r_0^2 M}
\ .\label{eqn:Lamdef}\end{equation}
The calculation in Sect.~\ref{sec:sol} supposes $\Lambda^2\ll 1$,
which holds when $h^2$ is sufficiently small, {\it i.e.\/}, when the preferential attraction is sufficiently weak.
This inequality condition is paraphrased more conveniently at the end of Sect.~\ref{sec:sol}.
\\

\begin{figure}
\includegraphics[width=6cm]{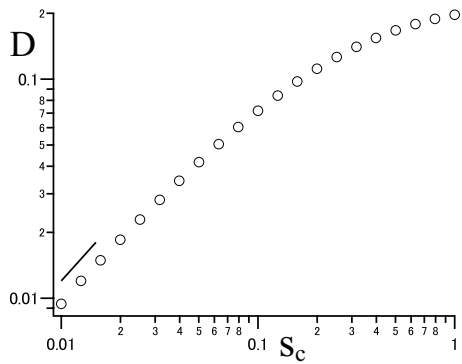}
\caption{\label{fig:zetde} Circles represent $D(s_{\rm c})$ calculated from
Eq.~(\ref{eqn:Dd}).  The slope of unity is
shown by the short line for reference.  }
\end{figure}

After the calculation shown in Sect.~\ref{sec:calc}, 
we can rewrite Eq.~(\ref{eqn:spring}) as
\begin{equation}
\left( m+{2\pi \rho r_0^3\over 3} \right) {d^2 \zeta \over dt^2}
 =-\left\{ m\omega_0^2 \left[ 1+ \lambda^2  
D(s_{\rm c})  \right] \right\} \zeta
\label{eqn:perpend}\end{equation}   
up to the order of $\varepsilon$.
This represents the small oscillation of the particle about the equilibrium point. 
 Here, $\lambda$ is a dimensionless parameter defined as $h\sqrt{6\pi r_0/(Mm)} /\omega_0$; 
$D$ is defined as
\begin{equation}
D(s_{\rm c})\equiv 2\kappa^2 \int_1^\infty ds\ {\Xi(s)^2\over s^6}\ , \label{eqn:Dd}\end{equation}
which is plotted in Fig.~\ref{fig:zetde} with the aid of the software Mathematica (Wolfram Research). 
 The second term in the parentheses on the left-hand side (lhs)
of Eq.~(\ref{eqn:perpend}) represents the induced mass, which equals half the mass of the displaced mixture,
as stated below Eq.~(\ref{eqn:equaltime}).   The term involving $\lambda^2$ on the right-hand side (rhs)
represents the additional restoring force due to the deformed adsorption layer.
We can derive $D(s_{\rm c})\approx s_{\rm c}$ for a small $s_{\rm c}$, as shown by Eq.~(\ref{eqn:Dapp}).
In agreement with this, the numerical results in Fig.~\ref{fig:zetde} for a small $s_{\rm c}$ have
the slope of about unity.  
\\

From Eq.~(\ref{eqn:perpend}), we apply the equipartition theorem
to obtain 
\begin{equation}
 \langle \zeta^2\rangle =k_{\rm B} T\left[ m\omega_0^2 +{6\pi h^2 r_0 \over M}  D(s_{\rm c})\right]^{-1}
\label{eqn:equi}\ , \end{equation}
which gives the mean square displacement of the particle.
The sum in the braces of Eq.~(\ref{eqn:equi}) is the same as the one in the brackets of Eq.~(\ref{eqn:perpend}).  
Comparing Eq.~(\ref{eqn:equi}) with the first equation of Eq.~(\ref{eqn:equaltime}), we find
that the effect of the deformed adsorption layer is represented by 
the second term in the braces of Eq.~(\ref{eqn:equi}),
which comes from the additional restoring force.  This term
reduces the mean square displacement, {\it i.e.\/}, it suppresses the fluctuation
amplitude.   The suppression effect is more marked
as $h^2$ and $\xi_{\rm c}$ are larger.
Then, the adsorption layer is
also remarkable in its thickness and amplitude, considering Eq.~(\ref{eqn:phizero}).   \\

\section{\label{sec:calc} Calculation procedure}
Here, we describe the calculation procedure leading to Eq.~(\ref{eqn:perpend}), with
some details being relegated to Appendix \ref{app:detail}.  
 A similar procedure is taken in Ref.~\onlinecite{pre}, where a fluid membrane
fluctuating around a plane is studied.  We have only to modify the procedure
to consider the spherical surface of the particle in the present problem.

\subsection{Pressure tensor} 
The reversible part of the pressure tensor is given by \cite{Onukibook, ofk}
\begin{equation}
\Pi=\left(p-f+\mu\varphi-{M\over 2}\left|\nabla\varphi\right|^2 \right)\bm{1}+M\nabla\varphi\nabla\varphi
\ ,\label{eqn:pidef}\end{equation}
where $\bm{ 1}$ denotes the isotropic tensor and $\mu$ is given by Eq.~(\ref{eqn:hatmudef}).
In the above, the part other than $p\bm{1}$ is derived from the first term of Eq.~(\ref{eqn:glw}), and
$\varphi f'-f$ gives the osmotic pressure \cite{deg}.
Because $\nabla \cdot \Pi$ equals
$\nabla p+ \varphi \nabla \mu$, we have Eq.~(\ref{eqn:3Ddyn}).
Let the local stress exerted on the particle by
the ambient mixture  
be denoted by $\bm{F}(\bm{r}_{\rm s}, t)$, where $\bm{r}_{\rm s}$ represents a point on the
particle surface.  We have
\begin{equation}
\bm{ F}(\bm{r}_{\rm s}, t)=- \Pi\cdot\bm{n}
+\nabla_\parallel f_{\rm s}-{2f_{\rm s} \over r_0} \bm{ n}
\ .\label{eqn:Fplus}\end{equation}
Here, $\nabla_\parallel$ implies the projection of $\nabla$
on the tangent plane and $1/r_0$ gives the mean curvature of the particle surface.
The last two terms above come from the stress due to 
the two-dimensional pressure $-f_{\rm s}$, as discussed in Appendix A of Ref.~\onlinecite{visc}.  
The tangential components of $\bm{ F}$ vanish;
the contribution from $M\nabla\varphi\nabla\varphi$ of Eq.~(\ref{eqn:pidef})
cancels with the tangential stress due to $f_{\rm s}$, as described in Appendix D
of Ref.~\onlinecite{visc}.  In fact, we use Eq.~(\ref{eqn:phisurface}) to
rewrite the rhs of Eq.~(\ref{eqn:Fplus}) as
\begin{equation}
\left( - {p}+ f+ {M\over 2}\vert \nabla \varphi\vert^2
- \mu\varphi
-{h^2 \over M}
-{2f_{\rm s}\over r_0} \right)\bm{ n}\ ,
\label{eqn:Fplus2}\end{equation}
which is evaluated immediately outside the boundary layer generated in
the limit of zero viscosity.  See the statement below Eq.~(\ref{eqn:3Ddyn}).

\subsection{Profile in the unperturbed state\label{sec:unp}}
We write $\partial_r$ for
the differentiation with respect to $r$, and $\partial_r^2$
for $\partial_r\partial_r$.
As argued for a mixture in contact with a flat wall \cite{Cahn},  Eq.~(\ref{eqn:hatmudef})
yields
\begin{equation}
f'(\varphi^{(0)})-{M\over r} \partial_r^2 r \varphi^{(0)}=\mu^{(0)}
\ ,\label{eqn:equalmu}
\end{equation}
while Eq.~(\ref{eqn:phisurface}) yields
\begin{equation}
M{\partial\over\partial r}\varphi^{(0)} =- h
\quad{\rm as}\ r\to r_0+ \ .\label{eqn:phisurface0}\end{equation}
Solving these equations, we arrive at Eq.~(\ref{eqn:phizero}). 
The equilibrium profile around a sphere was studied
under conditions other than those considered here \cite{indekeu, upton, liu, hanke}.   \\

\subsection{\label{sec:eps} Equations at the order of $\varepsilon$}
Suppose a point, $(r_0,\theta,\phi)$, on the particle surface in the unperturbed state (Fig.~\ref{fig:sphere}).
By the particle motion, its $z$-coordinate changes from $z_0\equiv r_0\cos{\theta}$ to
$z_0+\zeta$.  The slip boundary condition up to the order of $\varepsilon$ gives
\begin{equation}
\left.v_r^{(1)}\right|_{z_0}={d \zeta^{(1)}\over d t}\cos{\theta}
\ ,\label{eqn:slip}\end{equation}
where the subscript $z_0$ means that the term is evaluated at the point on the surface 
in the unperturbed state, $(r_0,\theta,\phi)$.
Equation (\ref{eqn:phisurface}) is rewritten as
\begin{equation}
-h=\left. M\bm{n}\cdot \nabla \varphi \right|_{z_0+\zeta}
\ ,\label{eqn:phisurface1}\end{equation}
where the subscript $z_0+\zeta$ means that the term is evaluated at the point on 
the surface of the shifted particle.
The unit normal vector $\bm{n}$ remains the same during the particle motion (Fig.~\ref{fig:sphere}).
We have
\begin{equation}
\left. \varphi \right|_{z_0+\zeta}=\left. \varphi^{(0)}\right|_{z_0+\zeta}+\left.\varepsilon
\varphi^{(1)}\right|_{z_0}+{\cal O}(\varepsilon^2)
\ ,\end{equation}
where ${\cal O}(\varepsilon^2)$ means the terms
whose quotient divided by $\varepsilon^2$ remains finite in the limit of $\varepsilon \to 0$. 
The first term on the rhs above is rewritten as
\begin{equation}
\varphi^{(0)}(r_0)+ \left. \zeta {\partial \over \partial z}\varphi^{(0)}(r)\right|_{z_0}\ .
\end{equation}
Considering Eq.~(\ref{eqn:phisurface0}), we use these three equations to obtain
\begin{equation}
\left.{\partial \over \partial r}\varphi^{(1)}\right|_{z_0}
=-\zeta^{(1)}{\varphi^{(0)}}''(r_0)\cos{\theta}\ .\label{eqn:phisurface2}
\end{equation}
Using Eqs.~(\ref{eqn:equalmu}), (\ref{eqn:phisurface0}), and (\ref{eqn:phisurface2}),
we rewrite Eq.~(\ref{eqn:Fplus2}) so that it is evaluated on the particle
surface in the unperturbed state.  Integrating the result over the surface, we obtain
up to the order of $\varepsilon$
\begin{eqnarray}
&&{\cal F}_z
= 2\varepsilon \pi r_0^2 \int_0^\pi d\theta\ \sin{\theta}\cos{\theta}
\left[ -p^{(1)}-\mu^{(1)}\varphi^{(0)}\right.
\nonumber\\ 
&&\left. + \left(\varphi^{(1)}-{h\over M}\zeta^{(1)}\cos{\theta}\right)
\left({2h\over r_0}+{M\over r_0} \partial_r^2 r{\varphi^{(0)}} \right)\right]
\ ,\label{eqn:efu} \end{eqnarray}
where the integrand is evaluated at $(r_0,\theta,\phi)$. 
 We define ${\cal F}_z^{(1)}$ so that
${\cal F}_z$ equals $\varepsilon{\cal F}_z^{(1)}$ up to the order of $\varepsilon$. \\

As in Eq.~(\ref{eqn:pexp10}),  we define the coefficient functions $Q_{10}$ and $G_{10}$ 
so that we have
\begin{eqnarray}
&&{\tilde \mu}^{(1)}(\bm{r},\omega)
=Q_{10}(r,\omega)Y_{10}(\theta) \nonumber\\ 
&& {\rm and}\ {\tilde \varphi}^{(1)}(\bm{r},\omega)
=G_{10}(r,\omega)Y_{10}(\theta)\ .
\label{eqn:muexp10}\end{eqnarray}
We use the vector spherical harmonics
\begin{eqnarray}
&&\bm{P}_{10}(\theta,\phi)=\bm{e}_rY_{10}\ 
{\rm and}\ \\
&&\bm{B}_{10}(\theta,\phi)={1\over \sqrt{2}} \bm{e}_\theta {\partial\over \partial\theta} Y_{10}(\theta)\ ,
\label{eqn:vecharm}
\end{eqnarray}
where $\bm{e}_r$ and $\bm{e}_\theta$ denote
the unit vectors tangent to the coordinate line of $r$ and curve of $\theta$, respectively.
As in Ref.~\onlinecite{ofk}, we can assume
\begin{equation}
{\tilde {\bm{v}} }^{(1)}(\bm{r},\omega)=R_{10}(r,\omega)\bm{P}_{10}(\theta,\phi)
+T_{10}(r,\omega)\bm{B}_{10}(\theta,\phi)
\ ,\label{eqn:vexp10}
\end{equation} 
where the functions $R_{10}$ and $T_{10}$ are introduced. \\

Below, we derive the equations to be satisfied by the coefficient functions.
Using Eqs.~(\ref{eqn:pexp10}) and (\ref{eqn:vexp10}) in
Eqs.~(\ref{eqn:incomp}) and (\ref{eqn:3Ddyn}), we obtain
Eqs.~(\ref{eqn:rntn10})--(\ref{eqn:pphi2}), which generate
\begin{equation}
{r^2\over 2} \partial_r^2 R_{10} + 2r \partial_r R_{10}={{\varphi^{(0)}}'Q_{10}\over i\rho\omega }
\ .\label{eqn:eqR10}\end{equation}
We introduce $\zeta_{10}$ so that we have
\begin{equation}
{\tilde \zeta}^{(1)}(\omega)=\sqrt{{3\over 4\pi}}\zeta_{10}(\omega)
\ .\end{equation}  
Equation (\ref{eqn:slip}) yields
\begin{equation}
\left.R_{10}\right|_{z_0}=-i\omega\zeta_{10}
\ ,\label{eqn:slip2}\end{equation}
and $R_{10}$ vanishes far from the particle, as stated below Eq.~(\ref{eqn:perexp}).
If $h$ vanishes, 
$\mu$ and $\varphi$ are constant; the constant solutions satisfy Eqs.~(\ref{eqn:hatmudef}), 
(\ref{eqn:phisurface}), (\ref{eqn:phidyn}), and (\ref{eqn:musurface}) together with
the boundary conditions far from the particle. Then, we find $R_{10}(r,\omega)=
-i\omega\zeta_{10} r_0^3/r^3$
from Eq.~(\ref{eqn:eqR10}) and the boundary conditions.
Thus, otherwise, $R_{10}$ is given by the sum of this solution for $h=0$ and
terms dependent on $h$. \\

Equation (\ref{eqn:phidyn}) yields
\begin{equation}
-i\omega G_{10}=-R_{10}{\varphi^{(0)}}'+L\left( \partial_r^2+{2\over r}\partial_r-{2\over r^2}
\right)Q_{10}
\ .\label{eqn:subst010}
\end{equation} 
If we assume $L$ to vanish from the beginning, Eq.~(\ref{eqn:subst010}) gives
\begin{equation}
i\omega G_{10}=R_{10} {\varphi^{(0)}}'\ ,\label{eqn:out}
\end{equation} which is differentiated with respect to $r$ 
to give ${\varphi^{(0)}}' \partial_r R_{10}=0$ at $r\to r_0+$ with the aid of Eq.~(\ref{eqn:phisurface2}).
Thus, then, we have $\partial_r R_{10}\to 0$ at $r\to r_0+$ for $h\ne 0$.
It is impossible to impose this boundary condition for any nonzero $h$ because 
Eq.~(\ref{eqn:eqR10}) has  
two other boundary conditions mentioned at and below Eq.~(\ref{eqn:slip2}).
This contradiction means that, if $h$ does not vanish,  Eq.~(\ref{eqn:out}) is valid only outside the 
boundary layer.
Then, even when $L$ approaches zero, the second term on the rhs of Eq.~(\ref{eqn:subst010}) cannot be 
neglected  inside the boundary layer because of
the steep change in $Q_{10}$.  
A similar problem occurs in the study of a fluid membrane fluctuating
around a plane \cite{pre}; 
the thickness of the boundary layer of the chemical potential vanishes
in the limit of $L\to 0+$. 
The solution outside the boundary layer is called the outer solution \cite{bender}.
Equation (\ref{eqn:out}) thus holds only if
$G_{10}$ and $R_{10}$ are replaced by their respective outer solutions. 
\\

\subsection{\label{sec:nond} Nondimensionalization}
We introduce the dimensionless fields
\begin{eqnarray}
&& {\cal G}(s,\omega)\equiv {MG_{10}(r,\omega)\over h\kappa\zeta_{10}(\omega)}\nonumber\\
&& {\rm and}\ {\cal Q}(s,\omega)\equiv {r_0^2 Q_{10}(r,\omega) \over h\zeta_{10}(\omega)}
\ ,\label{eqn:GQ}\end{eqnarray}
which do not diverge when $h$ vanishes because of the statement below Eq.~(\ref{eqn:slip2}). 
Hereafter, we sometimes write ${\cal G}$ or ${\cal G}(s)$ for
${\cal G}(s,\omega)$,
and ${\cal G}'$ for $\partial_s {\cal G}$.  These ways of writing
are also applied for other dimensionless functions.  
From Eq.~(\ref{eqn:phisurface2}) and the statement below Eq.~(\ref{eqn:perexp}), we have
\begin{equation}
{\cal G}'\to -1\ {\rm as}\ s\to 1+\ {\rm and}\quad {\cal G}\to 0\ {\rm as}\ s\to\infty\ .
\label{eqn:calGbound}\end{equation}
From Eq.~(\ref{eqn:musurface}) and the statement below Eq.~(\ref{eqn:perexp}), we have
\begin{equation}
{\cal Q}'\to 0\ {\rm as}\ s\to 1+\ {\rm and}\quad {\cal Q}\to 0\ {\rm as}\ s\to\infty\ .
\label{eqn:calQbound}\end{equation}
 We write ${\cal G}_{\rm out}$ and ${\cal Q}_{\rm out}$
for the outer solutions of ${\cal G}$ and ${\cal Q}$, respectively.
Equation (\ref{eqn:out})  gives the first of our key equations,
\begin{equation}
{\cal G}_{\rm out}=-\Xi {\cal R}_{\rm out}
\ ,\label{eqn:out2}\end{equation}
where ${\cal R}_{\rm out}$ denotes the outer solution of
\begin{equation}
{\cal R}(s, \omega)\equiv {iR_{10}(r,\omega)\over \omega\zeta_{10}(\omega)}
\ .\end{equation} 

\bigskip
We rewrite Eq.~(\ref{eqn:eqR10}) as 
\begin{equation}
\left({s^2\over 2} \partial_s^2 +2s \partial_s \right){\cal R}=\Lambda^2 \Xi {\cal Q}
\ .\label{eqn:calR}\end{equation}
Equation (\ref{eqn:slip2}) and the statement below it give
\begin{equation}
{\cal R}\to 1\ {\rm as}\ s\to 1+\ {\rm and}\quad {\cal R}\to 0\ {\rm as}\ s\to\infty\ .
\label{eqn:calRbound}\end{equation}
Applying the method of variation of parameters to Eqs.~(\ref{eqn:calR}) and
(\ref{eqn:calRbound}), we obtain the second of our key equations,
\begin{equation}
{\cal R}(s)=s^{-3}+{2\over 3}\Lambda^2 \int_1^\infty d\sigma
\Gamma_{\rm R}(s,\sigma) \Xi(\sigma){\cal Q}(\sigma)
\ ,\label{eqn:Rsol}\end{equation} 
where the kernel is defined by
\begin{equation}
\Gamma_{\rm R}(s,\sigma)=\left\{\begin{array}{ll}
s^{-3}\sigma^{-1}\left(1 - \sigma^3\right)     & {\rm for}\  \sigma <s \\
s^{-3}\sigma^{-1}\left (1- s^3\right)   & {\rm for}\  s\le\sigma 
\end{array}\right.
\ .\end{equation}

\bigskip
Picking up terms at the order of $\varepsilon$ of Eq.~(\ref{eqn:hatmudef}) yields
\begin{equation}
-\mu^{(1)}=\left( M\Delta -a \right) \varphi^{(1)}
\ .\label{eqn:mu1}\end{equation}
Substituting Eq.~(\ref{eqn:muexp10}) into the Fourier transform of Eq.~(\ref{eqn:mu1}), we obtain
\begin{equation}
-{\cal Q}=\kappa \left(\partial_s^2+{2\over s}\partial_s-{2\over s^2}-{1\over s_{\rm c}^2}
\right) {\cal G}\label{eqn:calG}
\end{equation}
with the aid of Eq.~(\ref{eqn:GQ}).
Applying the method of variation of parameters to Eq.~(\ref{eqn:calG}), 
we use the second condition of Eq.~(\ref{eqn:calGbound}) to obtain the last of our key equations,
\begin{equation}
{\cal G}(s)= {\cal G}'(1+) \Xi(s)
+{s_{\rm c}\over \kappa\sqrt{s}}\int_1^\infty d\sigma\ \sqrt{\sigma} \Gamma_{\rm G}(s,\sigma)
{\cal Q}(\sigma)\ ,\label{eqn:g10}
\end{equation}
where the kernel $\Gamma_{\rm G}$ is defined by Eq.~(\ref{eqn:kernel}) and
${\cal G}'(1+)$ denotes ${\cal G}'(s)$ at $s\to 1+$.
Because of the first condition of Eq.~(\ref{eqn:calGbound}),
we can substitute $-1$ into ${\cal G}'(1+)$ above. \\

We can define ${\cal Q}_{\rm out}(s,\omega)$ for $s>1$ in the limit of $L\to 0+$.
Subtracting it from ${\cal Q}(s, \omega)$, we define ${\cal Q}_{\rm in}(s,\omega)$ as the difference.
It vanishes outside the boundary layer, but cannot be neglected in the following integral.
Assuming that $s$ lies outside the boundary layer in Eq.~(\ref{eqn:g10}), we obtain
\begin{eqnarray}
&&{\cal G}_{\rm out}(s)=-\Xi(s)\left[ 1+ {1\over \kappa}\int_1^\infty d\sigma\ {\cal Q}_{\rm in}(\sigma)\right]
\nonumber\\
&&\quad+{s_{\rm c}\over \kappa\sqrt{s}}\int_1^\infty d\sigma\ \sqrt{\sigma} \Gamma_{\rm G}(s,\sigma)
{\cal Q}_{\rm out}(\sigma)
\label{eqn:g10out}
\end{eqnarray}
in the limit of $L\to 0+$.

\subsection{\label{sec:sol} Weak preferential attraction}
Assuming $\Lambda^2\ll 1$, we use the key equations in the preceding subsection
to calculate ${\cal F}_z$ in Eq.~(\ref{eqn:spring}).  To do so, 
we obtain ${\cal Q}$ up to the order of $\Lambda$
and use the result in Eq.~(\ref{eqn:Rsol}) to calculate ${\cal R}$ up to the order of $\Lambda^3$. 
To begin with, from Eqs.~(\ref{eqn:out2}) and (\ref{eqn:Rsol}), we derive 
\begin{equation}
{\cal G}_{\rm out}(s)=-{\Xi(s)\over s^3} +{\cal O}(\Lambda^2)
\ ,\label{eqn:goutapp}\end{equation}
which leads to
\begin{equation}
{\cal G}'_{\rm out}(1+)=-1-{3\over \kappa}+{\cal O}(\Lambda^2) 
\ .\label{eqn:partcalG}\end{equation}
Equation (\ref{eqn:calG}) remains valid if ${\cal Q}$ and ${\cal G}$ are respectively replaced by 
${\cal Q}_{\rm out}$ and ${\cal G}_{\rm out}$.  
Thus, we can do the same replacement in Eq.~(\ref{eqn:g10}) although
${\cal G}_{\rm out}'(1+)$ is given not by the first condition of Eq.~(\ref{eqn:calGbound}) but 
by Eq.~(\ref{eqn:partcalG}). 
Thus, in the limit of $L\to 0+$, we find \cite{com}
\begin{eqnarray}&&
{s_{\rm c}\over \kappa\sqrt{s}}\int_1^\infty d\sigma\ \sqrt{\sigma} \Gamma_{\rm G}(s,\sigma){\cal Q}_{\rm out}(\sigma)
\nonumber\\
&&\quad= -s^{-3}\Xi(s)+\left(1+{3\over \kappa}\right)\Xi(s) 
+{\cal O}(\Lambda^2)\ .
\label{eqn:g10out2}\end{eqnarray}
Substituting Eq.~(\ref{eqn:g10out2}) into Eq.~(\ref{eqn:g10out}),  we 
use Eq.~(\ref{eqn:goutapp}) to obtain 
\begin{equation}
\int_0^\infty d\sigma\ {\cal Q}_{\rm in}(\sigma)=3+{\cal O}(\Lambda^2)\ .
\label{eqn:Qin}\end{equation}
At Eq.~(\ref{eqn:qoutapp}), ${\cal Q}_{\rm out}$ is calculated up to the order of $\Lambda$.\\

Taking the limit of $s\to 1+$ in Eq.~(\ref{eqn:g10}), we use Eqs.~(\ref{eqn:g10out2}) and (\ref{eqn:Qin}) to find
\begin{equation}
{\cal G}(1+)= {1\over \kappa}+{\cal O}(\Lambda^2)
\end{equation}
with the aid of Eqs.~(\ref{eqn:Xidef}) and (\ref{eqn:kappadef}).
Thus, up to the order of $\Lambda$, the difference in the first parentheses of
Eq.~(\ref{eqn:efu}) vanishes.
We can rewrite the first two terms 
in the braces of Eq.~(\ref{eqn:efu}) by using Eqs.~(\ref{eqn:rntn10}) and (\ref{eqn:pphi2}).
Using this result and Eq.~(\ref{eqn:calRbound}),
we obtain
\begin{equation}
{\tilde {\cal F}}_z^{(1)}(\omega)=-{2\pi\rho\omega^2 r_0^3\over 3} \left[2+{\cal R}'(1+)
+{\cal O}(\Lambda^4) \right]{\tilde \zeta}^{(1)}(\omega)
\ .\label{eqn:calf}\end{equation}
As shown in Appendix \ref{app:detail}, we can utilize Eqs.~(\ref{eqn:Rsol}), (\ref{eqn:Qin}), and
(\ref{eqn:qoutapp}) to obtain
\begin{equation}
{\cal R}'(1+)=
-3+18\Lambda^2\kappa\int_1^\infty ds\ {\Xi(s)^2\over s^6}+{\cal O}(\Lambda^4)
\ .\label{eqn:rprime2}\end{equation}
From Eqs.~(\ref{eqn:spring}), (\ref{eqn:calf}), and (\ref{eqn:rprime2}), 
we can derive Eq.~(\ref{eqn:perpend}).  In its braces, the 
terms of $O(\lambda^4)$ are neglected. \\

We can calculate ${\varphi}^{(1)}$ approximately from
the first term on the rhs of  Eq.~(\ref{eqn:goutapp}).
Comparing the values of $\varphi$ at the two points on the $z$-axis immediately outside the particle,
we find that, if $h$ is positive, the value at $z=r_0+\zeta$ is larger (smaller) than the value at $z=-r_0+\zeta$ when $\zeta$
is positive (negative).  Thus, within the approximation, 
the local difference of the mass density of the disliked component subtracted from
the mass density of the preferred component is larger 
on the front side of the moving particle than on the rear side.
This situation is shown by Fig.~\ref{fig:twosphere}(b)
if we regard the gray scale as representing the deviation of the local difference from 
its value far from the particle.  \\

It is usual that the specific gravity of the particle is about unity.
In Eq.~(\ref{eqn:lange2}), the normal-mode frequency is thus about  $\sqrt{2/3}\omega_0$
because of $m_{\rm ind}\approx m/2$.  
The Fourier transform of Eq.~(\ref{eqn:perpend}) determines
the normal-mode frequency $\omega$ changed by the deformed adsorption layer.
Because the change is small in our perturbative calculation, 
we can use $\sqrt{2/3}\omega_0$ instead of $\omega$ in evaluating Eq.~(\ref{eqn:Lamdef}) approximately.     
As mentioned in the fourth paragraph of Sect.~\ref{sec:intro}, $s_{\rm c}=\xi_{\rm c}/r_0\ll 1$
is assumed  in our hydrodynamics, which leads to $\kappa\approx 1$ in Eq.~(\ref{eqn:kappadef}).
Thus, the condition $\Lambda^2\ll 1$ can be identified with $\lambda^2/3\ll 1$, where 
 $\lambda$ is defined above  Eq.~(\ref{eqn:Dd}). Thus, assuming $\Lambda^2\ll 1$ amounts to assuming a sufficiently weak
surface field.  We also find that our approximate calculation supposes a sufficiently large 
$\omega_0^2$.  This is reasonable, considering that the deformation of the adsorption layer becomes smaller
as the particle moves more slowly.  \\

\section{\label{sec:res} Discussion}
The deformed adsorption layer generates the additional 
restoring force of a trapped particle, as shown in Eq.~(\ref{eqn:perpend}), and thus
its mean square displacement is reduced, as shown by Eq.~(\ref{eqn:equi}).
No additional force can be
calculated if we regard the ambient mixture as a simple bath and disregard its
hydrodynamic effect, as shown in Appendix \ref{app:simple}.
The suppression effect of the adsorption layer on the fluctuation amplitude was also
pointed out for a fluid membrane immersed in a near-critical binary fluid mixture \cite{pre, preprint}.
The relevance of the hydrodynamic effect is less distinct around the membrane, however, 
because incorrect but nonzero additional force can be calculated for the membrane even without
the hydrodynamic effect being considered, as 
mentioned at the end of Appendix \ref{app:simple}.  \\

For example, suppose a mixture of nitroethane and 3-methylpentane, which has 
$T_{\rm c}=300$ K. The mixture has $\xi_{\rm c}\approx 10$ nm for $T-T_{\rm c}\approx 0.3$ K 
at the critical composition \cite{iwan}.  
On the basis of the discussion in Ref.~\onlinecite{liu},
$h\approx 10^{-6}$ m$^3$/s$^2$ is estimated in Ref.~\onlinecite{pre} 
for a glass surface and an organic mixture, between which  
a strong hydrogen bonding is formed.
Supposing a surface not forming the hydrogen bonding so strongly, 
we use $h=10^{-7}$ m$^3$/s$^2$.
The coefficient of the square gradient term in the free-energy density is sometimes called
the influence parameter \cite{rowl,poser}.  
This parameter can be defined in general for each pair of components,
$A$-$A$, $B$-$B$, and $A$-$B$, where $A$ and $B$ represent the kinds of
the two components.  The coefficient $M$ is for the last pair, which can be
 regarded roughly as the geometric mean of the parameters for the 
first two pairs \cite{aiche}.  
We cannot find out the data for the influence parameters
of the mixture mentioned above.  Judging from the data
for the pure fluid of alkane,
we use $M=10^{-16}$ m$^7$/(s$^2$kg) \cite{vdwexp}.  
For these parameter values, as discussed below Eq.~(\ref{eqn:bc}), 
a correlation length longer than about $10$ nm cannot be assumed in the regime of the Gaussian model.
\\

The spring constant of the optical tweezers typically ranges from $10^{-3}$ to $10^{-6}$ kg/s$^2$ \cite{daniel}.
Let us assume it to be $m\omega_0^2=4.14 \times 10^{-4}$ kg/s$^2$,
which leads to $\langle\zeta^2\rangle \approx 10$ nm$^2$ in Eq.~(\ref{eqn:equaltime})
without the ambient near-criticality being assumed.
Suppose a particle having $r_0=100$ nm with the specific gravity being about unity.
We have $s_{\rm c}=0.1$ for the value of $\xi_{\rm c}$ mentioned above.
As mentioned 
in the last paragraph of Sect.~\ref{sec:sol}, our 
formulation supposes $\xi_{\rm c} \ll r_0$, {\it i.e.\/}, $s_{\rm c}\ll 1$,
which is satisfied by the values above.
The prefactor of $D$ in Eq.~(\ref{eqn:equi})
is calculated as $6\pi h^2r_0/M \approx 2 \times 10^{-4}$ kg/s$^2$.
Thus, we have $\lambda^2\approx 0.48$, which satisfies the condition for
the weak preferential attraction, mentioned at the end of Sect.~\ref{sec:sol}.
According to Eq.~(\ref{eqn:equi}),  $\langle\zeta^2\rangle$
is reduced to about $9.5$ nm$^2$ by the deformed adsorption layer.  Its square root is much smaller than $r_0$,
which justifies the assumption of a small amplitude of the particle fluctuation.
The change in $\langle\zeta^2\rangle$ from $10$ to $9.5$ nm$^2$ 
can be expected to be measured by means of the recent experimental technique, mentioned in Sect.~\ref{sec:intro},
although the increased turbidity in the near-critical mixture may make the measurement  more delicate.  
The prefactor of $D$ contains $h^2/M$, while
Eq.~(\ref{eqn:phizero}) contains a factor $h/M$.
 Thus, we can determine the material constants,
$h$ and $M$, experimentally from the suppression effect and unperturbed profile if they are both measured.  
Similar discussions are found in Refs.~\onlinecite{pre} and \onlinecite{relax}.  \\

It is assumed in our calculation 
leading to the result, Eq.~(\ref{eqn:equi}), that a rigid sphere fluctuates with a small amplitude, that the mixture
in the homogeneous phase is not too close to the critical point, and
that the particle surface attracts one component weakly.
Thus, we cannot apply the result beyond 
the parameter range generating a weak suppression effect.
A larger suppression effect may be observed experimentally in a setup with some 
appropriate parameter values.
Predicting this requires future theoretical or numerical studies not requiring the assumptions above. 
The present study can be a guide for them. 
We also assume the preferential attraction to be caused by a short-range interaction and
to be represented only by the surface field.  How the suppression effect is altered
without these assumptions also remains to be studied.
We assume a particle size much larger than the correlation length to use the hydrodynamics
based on the coarse-grained free-energy functional.
For a smaller particle,  the critical concentration
fluctuation of the mixture would influence the particle motion, and
a procedure other than that used in the present study should be taken.   This is also a future problem. \\

\acknowledgements{ The author thanks K. Fukushima for the information regarding Ref.~\onlinecite{litim}.
Discussion with D. Goto is appreciated.}

\appendix
\section{Validity of the Gaussian model \label{app:renorm}}
In cases more general than considered in the Gaussian model, we can use the renormalized local functional theory,
which was proposed by Okamoto and Onuki \cite{JCP, PRE}.
They studied the universal properties of a near-critical binary fluid mixture between
two parallel walls or two spheres.   
Here, we assume that the mixture has the critical composition far from the solid surface.
The order parameter and correlation length are not homogeneous and are
respectively denoted by $\psi(\bm{r})\equiv \varphi(\bm{r})-\varphi_\infty$
and $\xi(\bm{r})$. The correlation length far from the surface, denoted by $\xi_\infty$,
is asymptotically equal to $\xi_0 \tau^{-\nu}$ as $\tau$ approaches zero, 
where $\xi_0$ is a nonuniversal constant, $\tau\equiv \left(T-T_{\rm c}\right)/T_{\rm c}$ is 
assumed to be positive, and the critical exponent $\nu$ is about $0.627$ for the binary fluid mixture.
We also use the critical exponents $\gamma\approx 1.239$ and $\eta\approx 0.024$. 
In the Gaussian model, $\xi_{\rm c}$ is regarded as $\xi_\infty$.  \\

Subtracting the volume integral of $\mu^{(0)}\varphi$ over $C^{\rm e}$
from Eq.~(\ref{eqn:glw}) with Eq.~(\ref{eqn:gauss}) gives the free-energy functional for the open system
in the Gaussian model.  Apart from an additional constant, 
the corresponding functional in the renormalized local functional theory
can be obtained by replacing $M$ and $f-\mu^{(0)}\varphi$
respectively by $C$ and $f_{\rm R}$ defined below. 
The coefficient $C$ depends on $\psi$ as
\begin{equation}
C(\psi) \equiv k_{\rm B}T_{\rm c} C_1w^{-\eta\nu}
\ ,\label{eqn:Cdef}\end{equation}
where $C_1$ is a nonuniversal constant and $w$ is defined by
$w\equiv \xi_0^{1/\nu}\xi^{-1/\nu}$, while $f_{\rm R}$ is given by
\begin{eqnarray}
&& f_{\rm R}(\psi)\equiv k_{\rm B}T_{\rm c} \left( {1\over 2} C_1\xi_0^{-2}w^{\gamma-1}\tau\psi^2 \right.\nonumber\\
&&\qquad\qquad \left. +
{1\over 4} C_1^2 u^*\xi_0^{-\epsilon}w^{\gamma-2\beta}\psi^4\right)\ ,
\label{eqn:f}\end{eqnarray}
as a result of the $\epsilon$-expansion.  In later numerical calculations,
we use $\epsilon=1$ and approximate the constant $u^*$ to be $2\pi^2/9$.  
Equation (\ref{eqn:f}) is found to be the same as Eq.~(3.5) of Ref.~\onlinecite{JCP} with the aid of
the scaling law $\left(\epsilon-2\eta\right)\nu=\gamma-2\beta$. 
Here, because the critical composition is assumed far from the surface, $\mu_\infty$ in 
Ref.~\onlinecite{JCP} vanishes.
\\

The free-energy functional after the replacement above is the one renormalized up to the local
correlation length without rescaling, as in the exact renormalization group theory \cite{litim}. 
Thus, we can apply the mean-field theory 
to calculate $\xi$ at each locus, which leads to
\begin{equation}
w=\tau+C_2w^{1-2\beta}\psi^2
\ ,\label{eqn:wtaupsi}\end{equation}   
where we use $C_2\equiv 3 u^* C_1 \xi_0^{2-\epsilon}$.
These equations can be found in Ref.~\onlinecite{JCP}; see 
Eqs.~(3.9) and (3.11) and the statement below its Eq.~(3.16) in this reference.  
Defining $U\equiv w/\tau$ and
\begin{equation}
s\equiv C_2U^{1-2\beta}\tau^{-2\beta}\psi^2
\ ,\label{eqn:henkan}\end{equation}
we rewrite Eq.~(\ref{eqn:wtaupsi}) as $U=1+s$.  Here, $s$ is defined unlike in the
text.  Equation (\ref{eqn:f}) is rewritten as 
\begin{equation}
f_{\rm R}(\psi)=
{k_BT_{\rm c} C_1\tau^{2\beta+\gamma}\over 2C_2\xi_0^2}U^{2\beta+\gamma-2}
\left( s+{s^2\over 6}\right)\ .
\label{eqn:fR}\end{equation}
When only values of $s$ much smaller than unity are significant, 
we can neglect $s^2/6$ above to use  
$C(\psi)\psi^2/(2\xi_\infty^2)$ as $f_{\rm R}(\psi)$ approximately.  Then, 
regarding $C$ and $\xi_\infty$ as $M$ and $\xi_{\rm c}$, respectively, 
we find that the Gaussian model is valid. \\

\begin{figure}
\includegraphics[width=6cm]{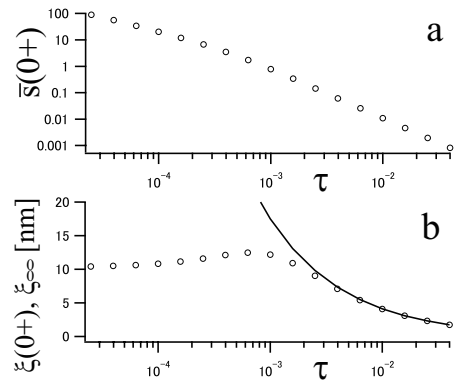}
\caption{\label{fig:szero} (a) We calculate ${\bar s}(0+)$, {\it i.e.\/}, $\lim_{z\to 0+} {\bar s}(z)$,
from Eqs.~(\ref{eqn:henkan}) and (\ref{eqn:bc}).
(b) Circles represent $\xi(0+)$, while the solid curve represents $\xi_\infty$. }
\end{figure}

By using the free-energy functional introduced above,
let us consider the range of $\xi_{\rm c}$ validating the Gaussian model.
To do so, limiting our discussion to cases of $\xi_{\rm c}\ll r_0$, we can assume
the mixture to occupy the semi-infinite space bounded  by
a flat surface.  Here, we take the $z$-axis so that the mixture lies in $z>0$.
The order parameter is not
rescaled in the renormalized local functional theory, and thus $f_{\rm s}$ remains unchanged
in the renormalization process.  
As discussed in Sect.~IIB of Ref.~\onlinecite{JCP}, the order-parameter profile at equilibrium
is obtained by minimizing the free-energy functional because it is renormalized up to
the local correlation length.
Writing ${\bar \psi}(z)$ for the profile, we obtain
\begin{equation}
 C({\bar \psi}(z))\left| {d{\bar \psi}(z)\over dz}\right|^2= 2 f_{\rm R}({\bar \psi}(z)) 
\label{eqn:bulk}\end{equation}
for $z>0$, and
\begin{equation}
C({\bar \psi(z)}) {d{\bar \psi}(z)\over dz}=-h
\label{eqn:bc}\end{equation}
as $z\to 0+$.  In the limit of $z\to \infty$, ${\bar \psi}(z)$ vanishes. 
Let us write ${\bar s}(z)$ for the variable $s$ determined by this profile ${\bar \psi}(z)$,
and ${\bar s}(0+)$ for ${\bar s}(z)$ in the limit of $z\to 0+$.
We can derive the differential equation with respect to ${\bar s}(z)$ from 
Eqs.~(\ref{eqn:henkan})--(\ref{eqn:bulk}), 
and derive the algebraic equation with respect to ${\bar s}(0+)$  from 
Eqs.~(\ref{eqn:fR})--(\ref{eqn:bc}).
The former shows that 
${\bar s}(z)$ decreases monotonically to zero as $z$ increases from zero to infinity \cite{preprint}.
Using the material constants mentioned in Sect.~\ref{sec:res}
and $\xi_0=0.23$ nm for the mixture \cite{iwan}, we numerically solve the latter to
obtain Fig.~\ref{fig:szero}(a).  We find that ${\bar s}(0+)$ is much smaller than unity
when $\tau$ is larger than about $2\times 10^{-3}$.  Then, we have ${\bar s}(z)\ll 1$ for any $z>0$
and the Gaussian model is valid.  Calculating the correlation length
immediately near the surface $\xi(0+)$ from ${\bar s}(0+)$, we plot the results in Fig.~\ref{fig:szero}(b),
where $\xi_\infty$ is also plotted for comparison.   These lengths turn out to agree only when the
 Gaussian model is valid, and then $\xi_\infty$ (or $\xi_{\rm c}$) is smaller than about $10$ nm.
As $\tau$ is smaller, $\xi(0+)$ reaches a plateau.
This is because the surface field prevents the mixture near the surface from approaching the
critical point.  The Gaussian model cannot describe the separation of $\xi(0+)$ and $\xi_\infty$. \\

Below, we briefly show that the procedure mentioned above properly yields
the well-known universal profile near the flat surface.
Details can be found in Ref.~\onlinecite{preprint}, where the undulation amplitude
of a fluid membrane is studied.
When $\tau$ decreases beyond the regime of the Gaussian model,
${\bar s}(0+)$ is much larger than unity.  Then,
we have ${\bar s}(z)\gg 1$ up to a positive $z$.   In this spatial region, 
Eqs.~(\ref{eqn:henkan})--(\ref{eqn:bulk}) lead to
${\bar \psi}' \approx -
\left( {\bar s}\tau \right)^{\beta+\nu} / \left( \sqrt{6C_2} \xi_0 \right)$
when $h$ is positive.  Noting $\left({\bar s}\tau\right)^\beta\approx \sqrt{C_2}{\bar\psi}$,  we find 
\begin{equation}
{\bar \psi}(z)\approx {1\over \sqrt{C_2} }\left[{\sqrt{6} \beta\xi_0\over \nu \left(z+l_0\right)}\right]^{\beta/\nu}
\ ,\label{eqn:jcppsi}\end{equation}
where $l_0\approx \sqrt{6}\beta \xi(0+)/\nu$ is found with the aid of Eq.~(\ref{eqn:bc}).
We find that ${\bar s}(z)\gg 1$ is valid for $z\ll \xi_\infty$. 
For $z\gg \xi_\infty$, the profile is shown to exhibit the same exponential decay as shown 
in the Gaussian model.
This is well known, as well as the profile of Eq.~(\ref{eqn:jcppsi}) \cite{rudjas, diehl94,smdilan}.  
When  $\tau$ is sufficiently small,  we have the region of $l_0\ll z\ll \xi_\infty$, where
Eq.~(\ref{eqn:jcppsi}) can be regarded as
\begin{equation}
{\bar \psi}(z)\propto \left( {\xi_0\over z}\right)^{\beta/\nu}  = \tau^\beta \left({z\over \xi_\infty}\right)^{-\beta/\nu}
\ ,\label{eqn:univ}\end{equation}
which is consistent with the universal form \cite{diehl97, diehl94, smdilan, smock}.
Equation (\ref{eqn:jcppsi})
 is the same as Eq.~(2.15) of Ref.~\onlinecite{JCP} in the case of $\tau=0$ and ${\bar \psi}(\infty)=0$ because of
the statement below Eq.~(3.15) in this reference.
In this case, the region of the exponential decay vanishes, and
the critical adsorption occurs.\\

Assuming $f_{\rm s}(\varphi)$ to be a linear function with a finite surface field, we can 
derive the universal form of the order-parameter profile, Eq.~(\ref{eqn:univ}), in terms of the
renormalized local functional theory, where the rescaling is not performed unlike in the
conventional renormalization group theory.
Here, we neglect the surface enhancement, {\it i.e.\/},  
the negative of the coefficient of the second-order term in $f_{\rm s}$.  This should
amount to assuming its bare value to be positive in the conventional renormalization group 
theory, where the surface enhancement is measured from the value at the special transition
occurring at $h=0$ (see Ref.~\onlinecite{diehl97} and Eq.~(3.95) of Ref.~\onlinecite{diehl86}).   
We think that nonzero surface enhancement should not  remarkably change
the result in the present study, where the preferential attraction is assumed to be
sufficiently weak.
\\

\section{Naive average based on Eq.~(\ref{eqn:glw}) \label{app:simple}}
Here, we calculate the mean square displacement by
regarding the mixture as a heat and particle bath.  This procedure is improper, but is shown for comparison
with the procedure described in the text.  
The following calculation is parallel with the one in the Appendix of  Ref.~\onlinecite{pre}.
We define $\Omega_{\rm b}$ as the first term of Eq.~(\ref{eqn:glw}) with $f$ replaced by $f-\mu^{(0)}\varphi$,
and define $\Omega_{\rm s}$ as the second term.  Their sum is denoted by
$\Omega$, which is the free-energy functional for the open system mentioned in the second paragraph of
Appendix \ref{app:renorm}.
Suppose that $\varphi$ and $\zeta$ deviate from $\varphi^{(0)}$ and zero, respectively.
Below, the resultant deviations of $\Omega_{\rm b}$, $\Omega_{\rm s}$, and $\Omega$, denoted by 
$\delta \Omega_{\rm b}$, $\delta\Omega_{\rm s}$, and
$\delta\Omega$, respectively, are calculated
up to the second order with respect to $\varphi_1\equiv \varphi-\varphi^{(0)}$ and $\zeta$.
We add the subscript $_\zeta$ to $C^{\rm e}$ and $\partial C$ to indicate 
their dependences on $\zeta$.  We find $\delta\Omega_{\rm s}[\varphi_1,\zeta]$ to be given by
\begin{eqnarray}
&&\int_{\partial C_\zeta}dS\ f_{\rm s}( \varphi )-\int_{\partial C_0}dS\ 
f_{\rm s}( \varphi^{(0)} )
\nonumber\\ &&
=-{h\zeta^2\over 2}\int_{\partial C_0}dS\
{\partial^2\over\partial z^2}\varphi^{(0)}
-h \int_{\partial C_\zeta}dS\ \varphi_1
\label{eqn:Omegas}\end{eqnarray}
with the aid of Eq.~(\ref{eqn:phisurface0}). Similarly, writing $\Phi^{(0)}$ for $\varphi^{(0)}-\varphi_\infty$, we find
$\delta\Omega_{\rm b}[\varphi_1,\zeta]$ to be given by
\begin{eqnarray}
&&\int_{C^{{\rm e}}_\zeta}d\bm{r}\ \left( a\Phi^{(0)}+M\nabla\varphi^{(0)}\cdot\nabla
\right)\varphi_1\nonumber\\ &&
-{\zeta^2\over 4} \int_{\partial C_0}dS\ \cos{\theta}{\partial\over\partial z}
\left( a{\Phi^{(0)}}^2+M\left|\nabla\varphi^{(0)}\right|^2\right)
\nonumber\\ &&
+{1\over 2}\int_{C^{\rm e}_0}d\bm{r}\ \left(a\varphi_1^2+M\left|\nabla\varphi_1\right|^2\right)
\ ,\label{eqn:Omegab}\end{eqnarray}
which can be rewritten with the aid of Eqs.~(\ref{eqn:equalmu}) and (\ref{eqn:phisurface0}).
 The sum of Eqs.~(\ref{eqn:Omegas}) and (\ref{eqn:Omegab}) gives $\delta\Omega[\varphi_1,\zeta]$, which is rewritten as
\begin{eqnarray}
&& {1 \over 2} \int_{\partial C_0}dS\ \left[\zeta^2 h {\varphi^{(0)}}''\cos^2{\theta}
-2\zeta M{\varphi^{(0)}}''\varphi_1\cos{\theta} \right.
\nonumber\\ &&\left.
-M \bm{n} \cdot \left(\varphi_1\nabla\varphi_1\right) \right]+
{1 \over 2} \int_{C^{\rm e}_0}d\bm{r}\ \left(a\varphi_1^2-M\varphi_1\Delta\varphi_1\right)\nonumber\\
\label{eqn:deltaOmega}\end{eqnarray}
with the aid of Eq.~(\ref{eqn:phisurface0}). 
Here, we also note that the surface integral of $(1-3\cos^2{\theta})$
over $\partial C_0$ vanishes.\\

In this procedure,
the probability density of $\varphi_1$ and $\zeta$ at equilibrium
is proportional to the Boltzmann weight,
$\exp{\left( -\delta \Omega^+/k_{\rm B}T\right) }$,
where $\delta\Omega^+$ is defined as $m\omega_0^2\zeta^2/2+\delta\Omega$.
Integrating this density with respect to
$\varphi_1$ yields the probability density of $\zeta$, from which we can find its variance, 
{\it i.e.\/}, the mean square displacement.  Apart from a multiplication constant,
we can obtain the result of this integration by replacing $\delta \Omega^+$  by
the minimum of $\delta \Omega^+$ for a given $\zeta$ in the Boltzmann factor
because the probability density considered here is a Gaussian distribution.
We write $\varphi_1^*$ for 
 $\varphi_1$ minimizing $\delta\Omega^+$, and thus $\delta\Omega$, for a given $\zeta$.
The stationary condition of Eq.~(\ref{eqn:deltaOmega}) with respect to $\varphi_1$ is found to be given by
Eq.~(\ref{eqn:mu1}) with its lhs being put equal to zero and Eq.~(\ref{eqn:phisurface2}) if
$\varphi^{(1)}$ and $\zeta^{(1)}$ are respectively replaced by $\varphi_1^*$ and $\zeta$. 
The condition is satisfied by
\begin{equation}
\varphi_1^*(\bm{r})=-\zeta{\varphi^{(0)}}'(r) \cos{\theta}\ . \label{eqn:shift}
\end{equation}
The time variable $t$ need not be specified in this procedure.
Replacing $\varphi_1$ with $\varphi_1^*$ in Eq.~(\ref{eqn:deltaOmega}), we find it to vanish.
This means that the probability density of $\zeta$ is not changed by the ambient mixture,
and that the mean square displacement 
remains the same as the first equation of Eq.~(\ref{eqn:equaltime}). 
In this procedure, considering that the rhs of Eq.~(\ref{eqn:shift})
equals $\varphi^{(0)} \left( r - \zeta\cos{\theta} \right) -\varphi^{(0)}(r)$
up to the order of $\zeta$,  the profile of $\varphi$ for a given $\zeta$ is 
obtained by the translational shift of the profile of Fig.~\ref{fig:twosphere}(a) and
such a deformed adsorption layer as shown in Fig.~\ref{fig:twosphere}(b) cannot be obtained.
It is thus reasonable that no additional force exerted on the particle
can be calculated by this improper procedure.  This is in contrast to the
corresponding result for a fluctuating fluid membrane. 
Even if the ambient mixture is regarded as a simple bath improperly, 
the adsorption layer around the membrane 
is deformed and nonzero additional force can be calculated although
the result is different from the
one obtained from the proper calculation with hydrodynamic consideration \cite{pre, preprint}. 

\section{Some details in the calculation \label{app:detail}}
Substituting 
Eq.~(\ref{eqn:vexp10}) into 
Eq.~(\ref{eqn:incomp}) gives
\begin{equation}
T_{10}={1\over\sqrt{2} r}\partial_r\left(r^2 R_{10}\right)
\ .\label{eqn:rntn10}\end{equation}
Picking up the terms with the order of $\varepsilon$ from 
the $r$- and $\theta$-components of Eq.~(\ref{eqn:3Ddyn})
and substituting
Eqs.~(\ref{eqn:pexp10}) and (\ref{eqn:vexp10})
into their Fourier transforms, we obtain
\begin{eqnarray}
&&-i\omega\rho R_{10}=-\partial_r P_{10}-\varphi^{(0)}\partial_rQ_{10}
\label{eqn:pphi}\\
&&{\rm and} \  {-i\omega\rho rT_{10}\over\sqrt{2}}=-P_{10}-\varphi^{(0)}Q_{10}
\ .\label{eqn:pphi2} \end{eqnarray}
These three equations give 
Eq.~(\ref{eqn:eqR10}). \\

Below, $I_{3/2}$ and $K_{3/2}$ 
represent modified Bessel functions.
Using ${\tilde s}\equiv 
s/s_{\rm c}$ and ${\tilde\sigma}\equiv 
\sigma/s_{\rm c}$, we define
\begin{equation}
\Gamma_{\rm Gc}( s,\sigma)\equiv
-{\hat w}(s_{\rm c})K_{3/2}({\tilde  s})K_{3/2}({\tilde \sigma})
{\tilde\sigma}\ .
\end{equation}
Here, we use
\begin{equation}
{\hat w}(s_{\rm c})\equiv {2e^{1/s_{\rm c}}\left[ 2s_{\rm c} {\rm cosh}s_{\rm c}^{-1} -\left(2s_{\rm c}^2+1\right)
{\rm sinh}s_{\rm c}^{-1}\right]\over \pi \left(1+s_{\rm c}\right) \kappa}
\ ,\end{equation}
which equals Eq.~(4.10) of Ref.~\onlinecite{ofk} if its $\zeta_{\rm c}$ is replaced by $s_{\rm c}$.
The kernel used in Eq.~(\ref{eqn:g10}) is defined as
\begin{equation}
\Gamma_{\rm G}( s,\sigma)
\equiv \left\{\begin{array}{ll} \Gamma_{\rm Gc}( s,\sigma)
+K_{3/2}({\tilde  s})I_{3/2}({\tilde \sigma})
{\tilde \sigma}& {\rm if}\ \sigma< s\\
\Gamma_{\rm Gc}( s,\sigma)
+I_{3/2}({\tilde  s})K_{3/2}({\tilde \sigma})
{\tilde \sigma}& {\rm if}\  s\le \sigma\end{array}\right.
\ ,\label{eqn:kernel}\end{equation}
which is the same as Eq.~(4.11) of Ref.~\onlinecite{ofk}.
In deriving Eq.~(\ref{eqn:g10}), we utilize the identity
\begin{equation}
K_{3/2}(s)I'_{3/2}(s)-I_{3/2}(s)K'_{3/2}(s)=s^{-1}\ ,\label{eqn:iden}
\end{equation}
and rewrite $I_{3/2}$ and $K_{3/2}$ in terms of the hyperbolic functions. \\ 

Equation (\ref{eqn:calG}) remains valid if ${\cal Q}$ and ${\cal G}$ are replaced by their respective outer solutions.
Substituting Eq.~(\ref{eqn:goutapp}) into the resultant equation gives
\begin{equation}
{\cal Q}_{\rm out}(s)=-{3\kappa\over s^2\Xi(s)} {d\over ds} {\Xi(s)^2 \over s^2}+{\cal O}(\Lambda^2)
\ .\label{eqn:qoutapp}\end{equation}
Here, we note that the rhs of Eq.~(\ref{eqn:calG}) with ${\cal G}(s)$ replaced by $\Xi(s)$ vanishes,
which is found by 
substituting Eq.~(\ref{eqn:gauss}) into Eq.~(\ref{eqn:equalmu})
and differentiating the result with respect to $r$.
Taking the limit of $s\to 1+$ of
the derivative of Eq.~(\ref{eqn:Rsol}) with respect to $s$,  we use Eqs.~(\ref{eqn:Xidef}), 
(\ref{eqn:kappadef}), and (\ref{eqn:Qin}) to obtain
\begin{equation}
{\cal R}'(1+)=-3-2\Lambda^2\left[-{3\over \kappa} +\int_1^\infty ds\ {\Xi(s)\over s} {\cal Q}_{\rm out}(s)\right]
\label{eqn:rprime}
\ .\end{equation}
Substituting Eq.~(\ref{eqn:qoutapp}) into Eq.~(\ref{eqn:rprime}), we obtain Eq.~(\ref{eqn:rprime2})
with the aid of the integration by parts.\\

Substituting Eq.~(\ref{eqn:Xidef}) into Eq.~(\ref{eqn:Dd}) yields
\begin{equation}
D(s_{\rm c})= {s_{\rm c}^3\over \left(1+s_{\rm c} \right)^{2}}
\left[{E_8(s_{\rm c})\over s_{\rm c}^2}+ {2E_9(s_{\rm c})\over s_{\rm c}}
+E_{10}(s_{\rm c})\right]
\ , \label{eqn:Ddef}\end{equation}
where we use
\begin{equation}
E_n(s_{\rm c})\equiv {2^ne^{2/s_{\rm c}}\over s_{\rm c}^n}\int_{2/s_{\rm c}}^\infty d\tau \ 
\tau^{-n}e^{-\tau} \ .\end{equation}
 As $s_{\rm c}\to 0+$, we have
\begin{equation}
E_n(s_{\rm c})\sim 1-{n\over 2}s_{\rm c}+{n(n+1)\over 4}s_{\rm c}^2+{\cal O}(s_{\rm c}^3)\ .
\label{eqn:Enapp}\end{equation} 
Equations (\ref{eqn:Ddef}) and (\ref{eqn:Enapp}) give the asymptotic relation
\begin{equation}
D(s_{\rm c})\sim s_{\rm c}-4s_{\rm c}^2+{\cal O}(s_{\rm c}^3)\quad {\rm as}\ s_{\rm c}\to 0+\ .
\label{eqn:Dapp}\end{equation}


\begin{thebibliography}{99}
\bibitem{const} Why the energy is determined uniquely is discussed around Eq.~(77) of Ref.~\onlinecite{pre}.
\bibitem{com} Substituting
Eq.~(\ref{eqn:qoutapp}) into the lhs of Eq.~(\ref{eqn:g10out2}) yields the rhs.
This is an alternative way of deriving Eq.~(\ref{eqn:g10out2}).
\end{thebibliography}
\end{document}